\documentclass[aps,twocolumn,amsmath,amssymb,nofootinbib, superscriptaddress, showpacs ,floatfix,prb]{revtex4}

\usepackage{latexsym}
\usepackage{graphicx}
\usepackage{times,psfrag,subfigure}
\usepackage{amsmath}
\usepackage{dcolumn}
\usepackage{bm, bbm}      
\usepackage{color}
\usepackage{latexsym,amsmath,amssymb,bm,euscript}
\def\re#1{(\ref{#1})}
\def\Tr{\mathop{\mathrm{Tr}}}
\def\sgn{\mathop{\textrm{sgn}}}

\newcommand{\beq}{\begin{equation}}
\newcommand{\eeq}{\end{equation}}
\newcommand{\beqarray}{\begin{eqnarray}}
\newcommand{\eeqarray}{\end{eqnarray}}
\newcommand{\Ref}[1]{Ref.~\onlinecite{#1}} 
\newcommand{\eq}[1]{Eq.~(\ref{#1})} 
\newcommand{\fig}[1]{Fig.~\ref{#1}} 

\begin{document}

\title{Andreev spectroscopy and surface density of states  for a three-dimensional time-reversal invariant topological superconductor}

\date{\today}

\author{Andreas P. Schnyder}
\email{a.schnyder@fkf.mpg.de}
\affiliation{Max-Planck-Institut f\"ur Festk\"orperforschung, Heisenbergstrasse 1, D-70569 Stuttgart, Germany}

\author{P. M. R. Brydon}
\email{brydon@theory.phy.tu-dresden.de}
\affiliation{Institut f\"ur Theoretische Physik, Technische Universit\"at
Dresden, 01062 Dresden, Germany}

\author{Dirk Manske}
\affiliation{Max-Planck-Institut f\"ur Festk\"orperforschung, Heisenbergstrasse 1, D-70569 Stuttgart, Germany}

\author{Carsten Timm}
\affiliation{Institut f\"ur Theoretische Physik, Technische Universit\"at Dresden, 01062 Dresden, Germany}

\begin{abstract}

A topological superconductor is a fully gapped superconductor that exhibits
exotic zero-energy Andreev surface states at interfaces with a normal metal.
In this paper we investigate the properties of a three-dimensional time
reversal invariant topological superconductor  
by means of a two-band model with unconventional pairing in both the inter-
and intraband channels. 
Due to the bulk-boundary correspondence the presence of Andreev surface states
in this system 
is directly related to the topological structure of the bulk wavefunctions,
which is characterized by a winding number. 
Using quasiclassical scattering theory we construct the
spectrum of the Andreev bound states that appear near the
surface and compute the surface density of states for various 
surface orientations.
Furthermore, we consider the effects of band splitting, i.e., the
breaking of an inversion-type symmetry, and demonstrate that in the absence of
band splitting there is a direct transition between the fully gapped
topologically trivial phase and the nontrivial phase, whereas in the presence
of band splitting there exists a finite region of a gapless nodal
superconducting phase between  
the fully gapped topologically trivial and nontrivial phases. 
\end{abstract}
\date{\today}


\maketitle

\section{Introduction}
\label{sec:sec1}

Due to the recent experimental discovery of the quantum spin Hall effect
\cite{kaneMelea08a,kaneMele08b,bernevig06,konig07} and the three-dimensional,
spin-orbit induced $\mathbbm{Z}_2$ topological insulator,\cite{fuKanePRL07,
fuKanePRB07, royPRB09, mooreBalentsPRB07, hsiehNature08}  there has
been a surge of interest in the study of topological insulating
electronic phases.  
Parallel to these developments, many workers have examined topological
superconductors,\cite{satoPRB06,schnyderPRB08,royArxiv08, qiZhangPRB08,
 schnyderPRL09,AIPconfProc,
 kitaev09,qiHughesPRL09,satoPRB09,satoFujiPRB09,hosur2010, qiZhangPRB10,
 schnyderNJP10} which are fully gapped unconventional superconductors that
exhibit exotic gapless Andreev surface states. 
Both topological insulators and topological superconductors can be described within a unified mathematical framework,\cite{schnyderPRB08, AIPconfProc, kitaev09, schnyderNJP10} which provides a complete
and exhaustive classification of topological phases of gapped free fermion systems in terms of discrete symmetries and spatial dimension. 
A distinctive property of these states is the bulk-boundary correspondence, which connects the presence of delocalized boundary modes to the topological structure of the bulk wavefunctions.

Notable examples of topological superconductors include the spinless  chiral
($p_x +ip_y$)-wave superconductor \cite{readGreen}
and  the B phase of superfluid ${}^3$He.\cite{schnyderPRB08,royArxiv08, qiHughesPRL09}
Here we focus on another type of topological superconductor,  which has been largely overlooked so far, namely the three-dimensional 
superconductor in symmetry class CI in the terminology adopted by Ref.~\onlinecite{schnyderPRB08} (following the Altland-Zirnbauer classification \cite{zirnbauerMathPhys96,altlandZirnbauer97}). 
The distinguishing characteristic of the CI topological superconductor is that, unlike any of the other three-dimensional topological states, it possesses a form of SU(2) spin- or pseudospin-rotation symmetry. In Ref.~\onlinecite{schnyderPRL09} a tight-binding model on the diamond lattice was proposed that realizes this nontrivial topological phase. 

In this paper we recast the model of Ref.~\onlinecite{schnyderPRL09}  into a form in which the topology of the Bogoliubov-de Gennes Hamiltonian  is completely determined by the phase structure of the superconducting gaps near the normal state Fermi surfaces. That is, we consider a two-band superconductor with exotic inter- and intraband gap functions, whose 
topological characteristics do not depend on the full Brillouin zone, but are controlled entirely by the properties of both the inter- and intraband gaps in the neighborhood of the Fermi surfaces. The reason for considering this case is two-fold: 
(i) It provides a clear  interpretation of the topological properties  in terms of the phase winding of the superconducting gap functions and
(ii) it allows for the straightforward application of the tools of quasiclassical scattering theory, a technique which has proven to be extremely useful for the study of the pairing symmetry in unconventional superconductors.\cite{bruderPRB90, hu1994, tanaka_kashiwaya95, kashiwaya_tanaka96, kashiwaya_tanaka00, eschrigIniotakisReview} Within this formalism, both the surface density of states and the spectrum of the Andreev bound states can be readily computed.  It is known that in some unconventional superconductors the presence of sub-gap surface bound states leads
to zero-energy anomalies in the surface density of states.\cite{tanaka_kashiwaya95, kashiwaya_tanaka96, kashiwaya_tanaka00, eschrigIniotakisReview} We will see that this rule also applies to the CI topological superconductor.

The remainder of the paper is organized as follows. Section~\ref{model_def} describes the model Hamiltonian and its symmetries. 
In Section~\ref{sec_winding} we introduce a bulk topological invariant and compute the phase diagram as a function of band width and
chemical potential. 
Section~\ref{sec_andreev} is concerned with the Andreev bound state spectrum 
and the surface density of states for various surface orientations.
We conclude with a summary and discussion in Section~\ref{conclusion}.

\section{Model Hamiltonian and Symmetries}
\label{model_def}

Our starting point is a time-reversal invariant  two-band  superconductor on a simple cubic lattice with inter- and intraband pairing, which has the form of a $4 \times 4$ Bogoliubov-de Gennes Hamiltonian. The mean-field Hamiltonian 
$\mathcal{H} = \sum_{\bm{k}} \Psi^{\ }_{\bm{k}} H ( \bm{k} ) \Psi^{\dag}_{\bm{k}} $ 
is diagonal in momentum space  with
\begin{subequations}  \label{def ham CI}
\begin{eqnarray}  \label{def ham CIa}
H ( \bm{k} ) 
=
\begin{pmatrix}
h ( \bm{k} ) & \delta ( \bm{k} ) \cr
\delta^{\dag} ( \bm{k} ) & - h^T ( - \bm{k} ) \cr
\end{pmatrix} 
\end{eqnarray}
and
$
\Psi_{\bm{k} }
=
(
a^{\ }_{\bm{k} \uparrow} , b^{\ }_{\bm{k} \uparrow} , a^{\dag}_{- \bm{k} \downarrow} , b^{\dag}_{- \bm{k} \downarrow} 
)^T ,
$
where $a_{\bm{k} \sigma }$ and $b_{\bm{k} \sigma }$ denote electron annihilation operators with spin $\sigma$ and momentum $\bm{k}$ for band one and two, respectively.
The normal state Hamiltonian $h(\bm{k})$ and the gap matrix $\delta ( \bm{k} )$ are given by
\begin{eqnarray} \label{nSmat}
h ( \bm{k} )
=
\begin{pmatrix}
\Theta_{1 \bm{k} } & 0 \cr
0 & \Theta_{2 \bm{k} } \cr
\end{pmatrix} 
\quad
\textrm{and}
\quad
\delta ( \bm{k} )
=
\begin{pmatrix}
\Delta_{\bm{k}} & \Phi_{\bm{k}} \cr
\Phi^{\ast}_{\bm{k}} & - \Delta_{\bm{k}} \cr
\end{pmatrix} ,
\end{eqnarray}
respectively.  
Here, the band dispersions are
\begin{eqnarray} \label{NSdispersion}
\Theta_{j \bm{k}}
&=&
t_j \left( \cos k_x + \cos k_y + \cos k_z \right) - \mu_j ,
\end{eqnarray}
with $j=1,2$, the intraband pairing  potential  is
\begin{eqnarray} \label{IntraPair}
\Delta_{\bm{k}}
&=&
\Delta_d \left( \cos k_x - \cos k_y \right) + \Delta_s ,
\end{eqnarray}
and the interband pairing potential takes the form \cite{footnote1}
\begin{eqnarray} \label{InterPair}
\Phi_{\bm{k}}
&=&
\Phi_0 \left( \sin k_x \sin k_y + i \sin k_z \right) .
\end{eqnarray}
\end{subequations}
We observe that Hamiltonian \re{def ham CI} is closely related to the model of
Ref.~\onlinecite{schnyderPRL09}. Specifically, by performing a particle-hole
transformation the tight-binding Hamiltonian analyzed in
Ref.~\onlinecite{schnyderPRL09} can be brought into a form in which
the momentum dependence of the gap functions along the Fermi surface have 
the same topology as the pairing potentials in Eq.~\re{def ham CIa}.
We note, however, that the present model is defined on a simple cubic lattice,
whereas in Ref.~\onlinecite{schnyderPRL09} a diamond lattice was considered. 
The energy spectrum of $H( \bm{k} )$ is composed of four bands with energies  $E ( \bm{k} ) 
\in
\left\{ -  \Lambda_{1 \bm{k}} , -  \Lambda_{2 \bm{k}} , +  \Lambda_{1 \bm{k}} ,  +  \Lambda_{2 \bm{k}}  \right\}$
and
\begin{eqnarray}
&&
\Lambda_{1 \bm{k} }
=
\sqrt{ \Delta^2_{\bm{k}} 
+
\frac{1}{4}  \left[   \Theta_{1 \bm{k}} - \Theta_{2 \bm{k}}  +  B_{\bm{k}}   \right]^2  } ,
\nonumber\\
&&
\Lambda_{2  \bm{k} }
=
\sqrt{ \Delta^2_{\bm{k}} 
+
\frac{1}{4}  \left[   \Theta_{1 \bm{k}} - \Theta_{2 \bm{k}} - B_{\bm{k}}   \right]^2  } ,
\end{eqnarray}
where
$
B_{\bm{k} }
=   \sqrt{ 4 \left|  \Phi_{\bm{k}}  \right|^2 + ( \Theta_{1 \bm{k}} + \Theta_{2 \bm{k}} )^2 } .
$

A topological superconductor belonging to the symmetry class CI  
satisfies two independent \emph{antiunitary} symmetries: time-reversal symmetry $\mathcal{T} = K \, U_T$, with $\mathcal{T}^2 = + 1$, and particle-hole symmetry $\mathcal{C} = K \, U_C$, with $\mathcal{C}^2 = - 1$. Here, $K$ denotes the complex conjugation operator. 
For the above example, time-reversal symmetry is expressed as
\begin{subequations}
\begin{eqnarray} \label{TRS}
U_T H^{\ast} ( - \bm{k} ) U^{\dag}_T = + H ( \bm{k} ), 
\end{eqnarray}
where $U_T = \mathbbm{1}$ is the $4 \times 4$ identity matrix.
The particle-hole symmetry 
can be expressed as \cite{footnotePHS}
\begin{eqnarray} \label{PHS}
U_C H^{\ast} ( - \bm{k} ) U^{\dag}_C 
=
- H ( \bm{k} ) ,
\end{eqnarray}
with $U_C = i \sigma_2 \otimes \sigma_0$. Here, and in the following, $\sigma_{1,2,3}$ denote the three Pauli matrices  and $\sigma_0$ is the $2 \times 2$ unit matrix.
By combining Eq.~\re{TRS} with Eq.~\re{PHS} we find that the gap matrix $\delta (\bm{k} ) $ is required to be
Hermitian, $\delta^{\dag} ( \bm{k} ) = \delta(\bm{k} )$.
Besides particle-hole and time-reversal symmetry, Hamiltonian~\re{def ham CI} satisfies another symmetry which 
is given by
\begin{eqnarray} \label{chiralS}
S^{\dag} H ( \bm{k} ) S 
= - H ( \bm{k} ),
\end{eqnarray}
\end{subequations}
with
$S = i U_T U_C =  - \sigma_2 \otimes \sigma_0$.
That is, $H ( \bm{k} )$ anti-commutes with the ``chiral'' symmetry operator $S$. 
The significance of symmetry \re{chiralS} is that it allows us to bring the Hamiltonian into
block off-diagonal form. Namely, we find that in the basis in which $S$ is diagonal,
$
\tilde{S} = W S W^{\dag} = \mathrm{diag} \left( - \sigma_0, + \sigma_0 \right)
$,
the Hamiltonian $H ( \bm{k} )$ takes the form
\begin{eqnarray} \label{off-diag}
\tilde{H} ( \bm{k} )
=
W H ( \bm{k} ) W^{\dag}
=
\begin{pmatrix}
0 & D ( \bm{k} ) \cr
D^{\dag} ( \bm{k} ) & 0 \cr
\end{pmatrix} ,
\end{eqnarray}
with the off-diagonal component 
\begin{eqnarray} \label{Doff}
D ( \bm{k} ) = h ( \bm{k} ) - i \delta ( \bm{k} )  ,
\end{eqnarray}
and the unitary transformation 
\begin{eqnarray} 
W
=
\frac{1}{\sqrt{2} }
\begin{pmatrix}
\sigma_0  & - i  \sigma_0  \cr
\sigma_0 & + i \sigma_0  \cr
\end{pmatrix} .
\end{eqnarray}
The block off-diagonal form \re{off-diag} is particularly useful to uncover the topological properties of the
ground state wavefunctions, as we will explain below.

\begin{figure}[t]
\begin{center} 
\includegraphics[width=0.48\textwidth,angle=-0]{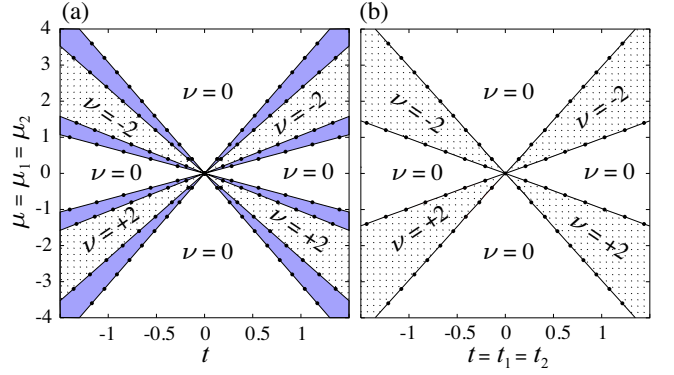}
\caption{
\label{fig:phaseDiags}
(color online).
Phase diagram for the CI topological superconductor, Eq.~\re{def ham CI}, as a function of bandwidth
and chemical potential with $\Phi_0 = 0.4$,  $\Delta_d = 0.4$, and $\Delta_s = 0$. 
The gapped phases are characterized by the even-numbered winding number $\nu$, Eq.~\re{Wno}.
Blue areas are bulk gapless phases.
(a) Inversion asymmetric case with split bands, $t_1 = t$, $t_2 = 0.9 t$ and  $\mu_1 = \mu_2$.
(b) Inversion symmetric case with degenerate bands, $t_1=t_2$ and $\mu_1 = \mu_2$.
}
\end{center}
\end{figure}

We end this Section by discussing the case of degenerate bands in Eq.~\re{def ham CI}, i.e., $\Theta ( \bm{k} ) \equiv \Theta_1 ( \bm{k})  =  \Theta_2 ( \bm{k} )$. This condition leads to an additional symmetry of $H(\bm{k})$, 
\begin{eqnarray} \label{invSym}
U_p H ( - \bm{k} ) U^{\dag}_p
= 
H (  \bm{k} ) ,
\end{eqnarray}
with the unitary matrix $U_p  =  \sigma_3 \otimes \sigma_2$. 
Eq.~\re{invSym} represents a type of inversion symmetry, as it is a $\emph{unitary}$ symmetry that
relates Bogoliubov-de Gennes Hamiltonians at $+\bm{k}$ and $-\bm{k}$ via a transformation that interchanges the two bands. 
With symmetry  \re{invSym} the energy spectrum becomes degenerate and
takes the simple form 
$E ( \bm{k} ) \in \left\{ - \Lambda_{\bm{k}}, + \Lambda_{\bm{k}} \right\}$ with
$\Lambda_{\bm{k}} = \sqrt{ \Theta^2_{\bm{k}} + \Delta_{\bm{k}}^2 + \left| \Phi_{\bm{k}} \right|^2 }$.

\section{Winding number and Phase Diagram}
\label{sec_winding}

To determine the topological properties of the model under consideration 
we first introduce an integer topological invariant, the winding number $\nu$.\cite{schnyderPRB08,AIPconfProc}
In order to do so it is convenient to
adiabatically deform $H(\bm{k})$ into a flat-band Hamiltonian. This can be achieved by means of a singular value decomposition.
First of all we note that for the off-diagonal block $D(\bm{k})$, Eq.~\re{Doff}, which is in general non-Hermitian, there exists a factorization of the form   $D(\bm{k})= U^{\dag} ( \bm{k} ) \Sigma ( \bm{k} ) V ( \bm{k} )$,
where $\Sigma ( \bm{k} )$ is a diagonal matrix with positive real numbers on the diagonal
and $U(\bm{k} ) $ and $ V ( \bm{k} ) $  are unitary matrices.  Direct calculation shows that the eigenvalues
of $\Sigma ( \bm{k} )$ are identical to the positive eigenvalues of the Bogoliubov-de Gennes Hamiltonian $H ( \bm{k} )$. 
For a fully gapped superconductor, it is possible to adiabatically deform the spectrum into flat bands with eigenvalues~ $+1$ and~$-1$. 
This procedure amounts to replacing $\Sigma ( \bm{k} ) $ by the unit matrix. Hence, the flat-band Hamiltonian $Q ( \bm{k} )$ in the off-diagonal basis reads
\begin{eqnarray}
Q ( \bm{k} )
=
\begin{pmatrix}
0 & q ( \bm{k} ) \cr
q^{\dag} ( \bm{k} ) & 0 \cr
\end{pmatrix} ,
\end{eqnarray}
with the unitary matrix $q ( \bm{k} ) = U^{\dag} ( \bm{k} ) V ( \bm{k} ) $.
In terms of the gap functions and band dispersions of model \re{def ham CI}, the off-diagonal block
of the flat-band Hamiltonian is given by
\begin{eqnarray}
q ( \bm{k} )
&=&
\begin{pmatrix}
\Lambda_+ B_{\bm{k}} - \Lambda_- \Theta_+    & - 2 i \Lambda_- \Phi_{\bm{k}} \cr
+2 i \Lambda_- \Phi^{\ast}_{\bm{k}}  &    \Lambda_+ B_{\bm{k}} + \Lambda_- \Theta_+  \cr
\end{pmatrix}
\frac{D ( \bm{k} )/2  }{  \Lambda_{1 \bm{k}} \Lambda_{2 \bm{k}} B_{\bm{k}} } ,
\nonumber\\
\end{eqnarray}
where $\Lambda_{\pm}  = \Lambda_{1 \bm{k} } \pm  \Lambda_{2  \bm{k} } $  and $\Theta_+ = \Theta_{1\bm{k} } + \Theta_{2 \bm{k}}  $ .
As a consequence of time-reversal invariance $q ( \bm{k} )$ satisfies $q^{T} ( - \bm{k} ) = q ( \bm{k} )$.
The topological invariant characterizing CI topological superconductors is defined as the winding number 
of the off-diagonal block $q ( \bm{k} )$,~\cite{schnyderPRB08, schnyderPRL09}
\begin{eqnarray} \label{Wno}
\nu 
=
\frac{1}{24 \pi^2 }
\int d^3 k \, \varepsilon^{\mu \nu \rho}
\Tr \left[ 
\left( q^{-1} \partial_\mu q \right) \left( q^{-1} \partial_\nu q \right) \left( q^{-1} \partial_\rho q \right)
\right] ,
\nonumber\\
\end{eqnarray}
where the integral is over the first  Brillouin zone. From the constraint $q^T ( - \bm{k} ) = q ( \bm{k} )$ it follows
that $\nu$ is even.

Next we use the topological invariant~\re{Wno} to analyze the phase diagram of $H (\bm{k})$, Eq.~\re{fig:phaseDiags}, as a function of chemical potential and band width. Fully gapped phases with different topological properties are separated by regions (or lines) of nodal superconducting phases (see Fig.~\ref{fig:phaseDiags}).
The condition for the existence of a gapless phase can be expressed in terms of a vanishing determinant, i.e., 
$\mathrm{det} \; H (\bm{k} ) = -  \left|  \mathrm{det} \; D  (\bm{k} ) \right|^2=0$. By use of Eq.~\re{nSmat} and Eq.~\re{Doff} we obtain
\begin{eqnarray} \label{gapClose}
\Delta^2_{\bm{k}}  + \left|  \Phi_{\bm{k}} \right|^2  + \Theta_{1\bm{k}} \Theta_{2\bm{k}}  = 0 ,
\quad
\Delta_{\bm{k}} \left( \Theta_{1 \bm{k}}  - \Theta_{2 \bm{k}} \right) = 0.
\end{eqnarray} 
Let us first focus on the inversion asymmetric case with split bands, $\Theta_{1 \bm{k}} \ne \Theta_{2 \bm{k}}$.  The above two conditions then reduce to
\begin{eqnarray} \label{condII}
\Delta_{\bm{k}} = 0, \quad   \left| \Phi_{\bm{k}} \right|^2 = - \Theta_{1 \bm{k} } \Theta_{2 \bm{k}} .
\end{eqnarray}
Provided $\sgn \Theta_{1 \bm{k}} = - \sgn \Theta_{2 \bm{k}}$,  Eqs.~\re{condII} have solutions describing nodal rings that appear
in the gapless phase of Fig.~\ref{fig:phaseDiags}(a)  (blue shaded area). These gap-closing lines in momentum space are topologically stable and 
are characterized by an integer topological charge,\cite{beriPRB2010} akin to the band touching points in graphene.
That is, the appearance of these nodal lines is generic and stable against (small) perturbations of the Hamiltonian, such as, e.g., the 
inclusion of higher $d$-wave gap harmonics in the intraband pairing potential.

In the presence of inversion symmetry \re{invSym}, with $\Theta_{\bm{k}} \equiv \Theta_{1 \bm{k}} = \Theta_{2 \bm{k}} $, the gap closing condition \re{gapClose} becomes
$\Delta^2_{\bm{k}}  + \left|  \Phi_{\bm{k}} \right|^2  + \Theta^2_{\bm{k}}  = 0$. Hence, there are in general four conditions that need to be satisfied for the gap to be zero,
which exceeds the number of free parameters $(k_x, k_y, k_z)$. In other words, if we consider tuning a single parameter, e.g., the band width $t$, to  drive a transition
from a topologically nontrivial phase to a topologically trivial phase, a gap closing can only occur at isolated points in the $(\bm{k}, t)$ parameter space [see Fig.~\ref{fig:phaseDiags}(b)]. Thus, the presence of inversion symmetry~\re{invSym} leads to a direct quantum phase transition between two distinct gapped phases (i.e, there is no intervening gapless phase).
The phase boundaries in the $( t , \mu)$-plane are given by $\mu = \pm  t$ and $\mu = \pm 3 t$, as shown in Fig.~\ref{fig:phaseDiags}(b). The fact that direct transitions from one gapped phase to another are only possible in the presence 
of inversion symmetry is a feature which is common
to three-dimensional topological phases.\cite{beriPRB2010} In particular, it also occurs in $\mathbbm{Z}_2$ topological insulators.\cite{murakamiPRB2008}

To determine the topological nature of the eight gapped phases in Fig.~\ref{fig:phaseDiags} (white and dotted areas), we computed the winding number $\nu$ numerically, by discretizing the integral~\re{Wno} over the Brillouin zone. It turns out that four phases are topologically nontrivial with winding number $\nu = \pm 2$. 
In these nontrivial phases, there appear linearly dispersing, mid-gap surface states when the system is placed next to a normal metal or insulating state. These exotic  Andreev bound states are robust against localization from random impurities.  We will study these surface states  in more detail in Section~\ref{sec_andreev}.
In passing, we note that for the inversion symmetric case of model~\re{def ham CI}, there exists an intimate connection between the topological properties of the Bogoliubov-de Gennes wavefunctions, as characterized by the  winding number $\nu$, and the Fermi surface topology in the normal state. 
In particular, the transition from a topologically trivial to a nontrivial phase in Fig.~\ref{fig:phaseDiags}(b)
 coincides with a change in Fermi surface topology.
A similar relationship has been previously reported for fully gapped time-reversal invariant spin-triplet superconductors.\cite{satoPRB09}

\begin{figure}[t]
\begin{center} 
\includegraphics[width=\columnwidth, angle=-0]{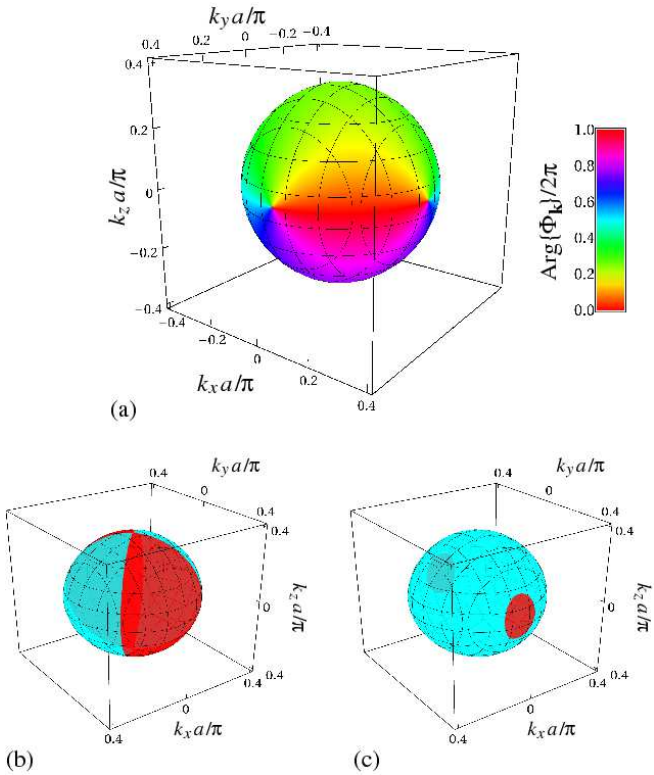}
\caption{
\label{fig:phases_of_gaps}
(color online). Panel (a): 
Variation of  $ \mathrm{arg} (\Phi_{\bm{k}} )$ over the Fermi surface. The argument of  $\Phi_{\bm{k}}$ shows four distinct singularities on the Fermi surface. The locations of these singularities are related by a four-fold rotational symmetry about the $z$ axis.
Panel (b) and (c):
Variation of  $\mathrm{sign} ( \Delta_{\bm{k}} )$ over the Fermi
surface.  Red indicates $\mathrm{sign} ( \Delta_{\bm{k}} ) = + 1$,
while blue is $\mathrm{sign} ( \Delta_{\bm{k}}) = -1$. The ratio between $s$- and $d$- wave components
is $\Delta_s / \Delta_d = 0$ and $\Delta_s / \Delta_d = 3/7$ in panel (b) and (c), respectively. 
The normal state band structure parameters are  $t_1 = t_2 = 1$ and $\mu_1 = \mu_2 = 1.75$.
}
\end{center}
\end{figure}

\subsection{Fermi surface topological invariant}

An important property of Hamiltonian \re{def ham CI} is that in the presence of inversion symmetry its topological characteristics are completely determined by the momentum dependence of the superconducting gap functions along the normal-state Fermi surface. To demonstrate this, 
we give an illustration of the topological properties of $H ( \bm{k})$ in terms of the phase structure of the gap functions 
on the Fermi surface. 

For definitiveness we consider a Fermi surface of spherical topology which is centered around the
$\Gamma$ point, $\bm{k}=0$, i.e. we focus on the region $t < \mu < 3 t$ in the $(t, \mu)$-plane of Fig.~\ref{fig:phaseDiags}(b). 
In the following we will hold the band structure parameters $t$ and $\mu$ fixed and use the ratio between $s$- and $d$-wave components in the intraband pairing \re{IntraPair} to tune the system form a topologically trivial to a nontrivial phase. 
Projected onto the Fermi surface, we find that $\Phi ( \bm{k} )$ has four first-order zeroes at momenta 
$\bm{k}_{1\pm} = ( \pm k_F, 0,0)$
and 
$\bm{k}_{2 \pm} = (0, \pm k_F, 0)$, with associated singularities in $\mathrm{arg} ( \Phi_{ \bm{k} } ) $ [see Fig.~\ref{fig:phases_of_gaps}(a)].
In other words, the real vector field $\Phi_{\bm{k}}$ over the Fermi surface exhibits
vortices at $\bm{k}_{1 \pm}$ and anti-vortices at $\bm{k}_{2 \pm}$ with winding number $+1$ and $-1$, respectively. 
The appearance of vortices in  $\Phi ( \bm{k} )$ is a necessary but not sufficient condition for the nontriviality of the model.
What is required in addition 
is that $\Delta ( \bm{k} )$ reverses sign between vortices of $ \Phi ( \bm{k} ) $ with opposite winding number. Figs.~\ref{fig:phases_of_gaps}(b) and \ref{fig:phases_of_gaps}(c) display the variation of 
the sign of $\Delta ( \bm{k} )$ over the Fermi surface for two different parameter choices, which both lead to a nontrivial state.
For  $ \Delta_d\cos(k_F) >  \Delta_s$ the sign of $\Delta_{\bm{k}}$ reverses
between $\bm{k}_{1\pm}$ and $\bm{k}_{2\pm}$, and the 
system is in the topologically nontrivial phase. If $\Delta_d\cos(k_F) <
\Delta_s$, on the other hand, the sign of $\Delta_{\bm{k}}$ is the same across
the Fermi surface, and so we have a topologically trivial state.

\section{Andreev bound states and surface density of states}
\label{sec_andreev}

A physical consequence of the nonzero winding number $\nu$ is the appearance of gapless Andreev bound states
at the surface of a CI topological superconductor or at an interface between a normal metal and a CI topological superconductor. The bulk-boundary correspondence relates the number of Andreev bound states to the topological number $\nu$. 
In this Section we derive the energy spectrum of the bound states and the surface density of states using quasiclassical scattering theory.
For simplicity, we focus on the inversion symmetric case of model \re{def ham CI}
and assume a spherically symmetric Fermi surface. But 
the results we obtain are expected to remain qualitatively unchanged upon inclusion of 
anisotropic Fermi velocities or  inversion asymmetric  perturbations.

\begin{figure}[t!]
\begin{center} 
\includegraphics[width=0.47\textwidth,clip]{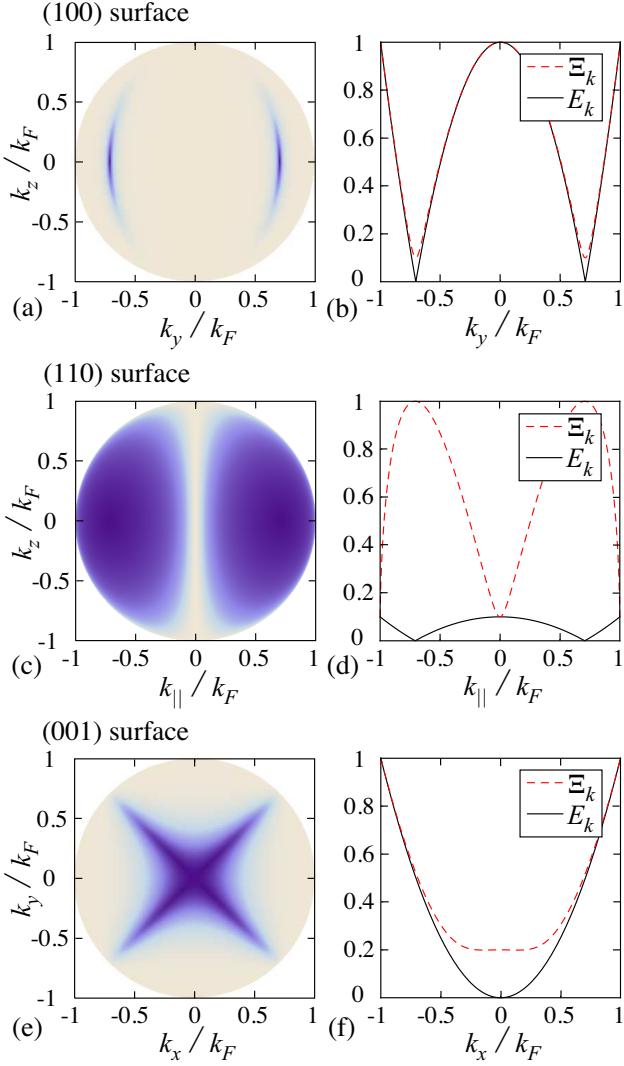}
\caption{
\label{fig:boundState}
(color online).
Panels (a), (c), and (e):
Energy of the upper branch of the interface states $E_{\bm{k}}$ [see Eq.~\re{eqForE}] as a fraction of
the gap amplitude $\Xi_{\bm{k}}$ for (a) the (100) interface perpendicular to  $\hat{\bm{e}}_x$, 
(c) the (110) interface perpendicular to $\frac{1}{\sqrt{2} } \left(  \hat{\bm{e}}_x   +  \hat{\bm{e}}_y  \right)$,
and (e) the (001) interface  perpendicular to $\hat{\bm{e}}_z$.
The color scale is such that blue corresponds to $E_{\bm{k}} / \Xi_{\bm{k}} =  0$, while grey is $E_{\bm{k}} / \Xi_{\bm{k}} =  1$.
Panels (b), (d), and (f): Comparison of the Andreev state dispersion $E_{\bm{k}}$ and the bulk gap $\Xi_{\bm{k}}$
for the same interface configurations as in panels (a), (c), and (e), respectively.
In all panels we take $\Delta_d = 1$, $\Delta_s = 0$ and $\Phi_0 = 0.2$.
}
\end{center}
\end{figure}

As we have seen in Section~\ref{sec_winding} the topological characteristics of the inversion symmetric model  \re{def ham CI} are fully determined by the phase structure of the pairing functions on the Fermi surface. This implies that we can capture the key topological
structure of the superconducting state by adopting an effective low-energy
quasiclassical description as long as the gap functions $\Delta ( \bm{r} )$
and $\Phi ( \bm{r} )$ are slowly varying over length scales of the
order of the inverse Fermi momentum  ${k}_F^{-1}$. 
Hence, we proceed by approximating the momentum dependence of
the gap functions in the vicinity of the Fermi surface by
\beqarray
\Delta_{\bm k} & = & \Delta[\lambda(k_{x}^2 - k_{y}^2)/k_{F}^2 + (1-\lambda)] , \\
{\Phi}_{\bm k} & = & \Phi(k_{x}k_y/k_{F}^2 + ik_z/k_F) \, = \, |\Phi_{\bm{k}}|e^{i\varphi_{\bm k}} .
\eeqarray
Here we have introduced the parameter $\lambda$ to tune $\Delta_{\bm k}$ from
$d_{x^2-y^2}$-wave symmetry ($\lambda=1$, topologically nontrivial) to
$s$-wave symmetry ($\lambda=0$, topologically trivial).

We can now apply standard methods to obtain the surface bound
states.~\cite{hu1994} 
We describe the system in terms of coordinates parallel ($\bm{r}_{\parallel}$)
and normal ($x_{\perp}$) to the interface. We assume that the superconductor
occupies the region defined by $x_{\perp} > 0$. 
We solve the Andreev equations using the ansatz
\beq
\Psi ({\bm k}_{\parallel}, {\bm r} ) = \sum_{j=1}^{4}\alpha_{j} \Psi_{j} (
{\bm k}_{\parallel}, {\bm r} )
\eeq
for the wavefunctions of a bound state of energy $E$, 
where the spinors are written
\begin{subequations} \label{eq_andreev_bound}
\begin{eqnarray} 
\Psi_{1} ( \bm{k}_{\parallel}, \bm{r} ) & = & \left(\begin{array}{cccc}
1,& 0,& u_{{\bm k}},& v_{\bm k}
\end{array} \right)^{T}e^{i{\bm k}\cdot{\bm r}}e^{-\kappa_{\bm k}x_{\perp}}
\\
\Psi_{2} ( \bm{k}_{\parallel}, \bm{r} ) & = & \left(\begin{array}{cccc}
1,& 0,& u_{\tilde{\bm k}},& v_{\tilde{\bm k}}
\end{array} \right)^{T}e^{i\tilde{\bm k}\cdot{\bm r}}e^{-\kappa_{\tilde{\bm k}}x_{\perp}}
\\
\Psi_{3} ( \bm{k}_{\parallel}, \bm{r} ) & = & \left(\begin{array}{cccc}
0,& 1,& v_{\bm k},& -u_{\bm k}
\end{array} \right)^{T}e^{i{\bm k}\cdot{\bm r}}e^{-\kappa_{\bm k}x_{\perp}}
\\
\Psi_{4} ( \bm{k}_{\parallel}, \bm{r} ) & = & \left(\begin{array}{cccc}
0,& 1,& v_{\tilde{\bm k}},& -u_{\tilde{\bm k}}
\end{array} \right)^{T}e^{i\tilde{\bm k}\cdot{\bm r}}e^{-\kappa_{\tilde{\bm k}}x_{\perp}}
\end{eqnarray}
\end{subequations}
with
\begin{subequations}
\beqarray
u_{\bm p} &=& \frac{\Delta_{\bm p}}{\Xi^2_{\bm p}}\left[E -
i\frac{p_{\perp}}{|p_{\perp}|}\sqrt{\Xi^2_{\bm p}-E^2}\right]\, , \\
v_{\bm p} &=& \frac{\Phi_{\bm p}}{\Xi^2_{\bm p}}\left[E +
i\frac{p_{\perp}}{|p_{\perp}|}\sqrt{\Xi^2_{\bm p}-E^2}\right]\, , \\
\kappa_{\bm p} 
& = & 
\frac{1}{ \left| v_F^{x_{\perp}} \right| }
\sqrt{\Xi_{\bm p}^2 - E^2}
\, ,
\eeqarray
\end{subequations}
and $\bm{v}_F =   \partial \Theta_{\bm{p}}  / \partial \bm{p} $ denotes the Fermi velocity.
We define the wavevectors ${\bm k}=({\bm k}_{\parallel},k_\perp)$ and $\tilde{\bm{k}} = ({\bm k}_\parallel,-k_\perp)$ with the requirement that $|{\bm k}| =
|\tilde{\bm k}| = k_F$.

The energy of the bound states is obtained by the condition that the equation
$\Psi( {\bm k}_\parallel, {\bm r})|_{x_\perp=0} = 0$ has a nontrivial solution
for the coefficients $\alpha_{j}$.
Although in general this yields a rather complicated expression for the bound 
state  energies, we can simplify matters considerably if we assume that
$\left| \Delta_{\bm{k}} \right| = \left| \Delta_{\tilde{\bm{k}}} \right|$ and 
$\left| \Phi_{\bm{k}} \right| = \left| \Phi_{\tilde{\bm{k}}} \right|$, which
holds for certain high-symmetry reflection planes. We hence find the bound
state energies 
\begin{eqnarray} \label{eqForE}
E_{{\bm k}_{\parallel}} ^2
&=& 
\frac{1}{2}
\left[
\Xi^2_{\bm k}  + \Delta_{\bm k} \Delta_{\tilde{\bm k}}  +  \left| \Phi_{\bm k} \right|^2  \cos ( \varphi_{\bm k}  - \varphi_{\tilde{\bm k}}  )
\right] .
\end{eqnarray}
This expression is valid in both the topologically trivial and
nontrivial cases. A zero-energy state is possible whenever the following two
conditions are both satisfied
\begin{eqnarray} \label{eq:conditions}
& \textrm{(i)} &
\quad
\left| \Phi_{\bm{k} } \right|  = 0  
\quad \textrm{or} \quad
 \cos ( \varphi_{\bm{k}} - \varphi_{\tilde{\bm{k}} } ) = -1   ,
\nonumber\\
& \textrm{(ii)} &
\quad 
\left| \Delta_{\bm{k}} \right| = 0 
\quad \textrm{or} \quad
\sgn \Delta_{\bm{k}} \sgn \Delta_{\tilde{\bm{k}}} = -1  .
\end{eqnarray}
We shall illustrate different possible combinations of
these conditions for the
appearance of the zero-energy states by examining three distinct cases: the
$(100)$ surface, the $(110)$ surface, and the $(001)$ 
surface. 

The surface bound states in unconventional superconductors can be
observed by scanning tunneling spectroscopy of the surface density of states 
(SDOS). It is therefore interesting to consider the SDOS for our three
surfaces, in order to determine the experimentally relevant signatures of the
topologically nontrivial phase.
To obtain the SDOS we must first calculate the quasiclassical
retarded Green's  
function $\mathcal{G}^{r}_{\bm{k}_{\parallel}}({\bm r},{\bm r}';E)$. This will
not be explicitly constructed here as it is rather laborious; for a detailed
discussion see~\Ref{kashiwaya_tanaka00} and references 
therein. Because we are dealing with a two-band system, the quasiclassical
Green's function $\mathcal{G}$ will be a $4 \times 4$ matrix in the Nambu-band
space. The SDOS is simply the local density of states (DOS) at the surface of the superconductor,
where the local DOS at the point ${\bm r}$ is defined as
\begin{eqnarray}
\rho ( E, {\bm r} )
=
- \frac{1}{\pi}
\sum_{\bm{k}_{\parallel} }
\textrm{Im} \left\{
\mathcal{G}^{r,11}_{\bm{k}_{\parallel}}
( {\bm r}, {\bm r}; \widetilde{E} )
+
\mathcal{G}^{r,22}_{\bm{k}_{\parallel}}
( {\bm r}, {\bm r}; \widetilde{E} )
\right\} .
\nonumber\\
\end{eqnarray}
Here $\mathcal{G}^{r,11}$ and $\mathcal{G}^{r,22}$ are the electron-like
Green's functions for bands one and two, respectively, and $\widetilde{E} = E + i
\Gamma$ contains the phenomenological broadening parameter $\Gamma$. In all our
calculations we set $\Gamma=0.01\Delta$.

\subsection{$(100)$ surface}

We consider first the appearance of zero-energy states at the $(100)$
surface for $\lambda=1$. From the conditions~\eq{eq:conditions} on the
intraband pairing we have $\text{sgn}\Delta_{\bm{k}}\text{sgn}\Delta_{\widetilde{\bm k}}=1$ for all ${\bm k}$, but
$\Delta_{\bm k}=0$ along the lines defined by ${\bm k}=(k_{x},\pm 
k_{x},k_{z})$. The interband potential is always non-zero for these momenta,
but we do have   
$\cos(\varphi_{\bm k} - \varphi_{\widetilde{\bm k}})=-1$ (i.e. a sign reversal
of $\Phi_{\bm k}$ upon reflection) whenever $k_z=0$. We hence
obtain two zero-energy states at 
${\bm k}_1 = (k_{F}/\sqrt{2},k_{F}/\sqrt{2},0)$ and 
${\bm k}_2 = (k_{F}/\sqrt{2},-k_F/\sqrt{2},0)$. These arguments still hold if we decrease
$\lambda$ towards 0.5, but the momenta ${\bm k}_1$ and ${\bm k}_2$ will move
towards one another, eventually merging at $\lambda=0.5$ where we have a
gapless state.

In~\fig{fig:boundState}(a) we plot the $\lambda=1$ surface states in units of $\Xi_{\bm k}$. The energy of the surface states only deviates
significantly from the bulk gap close to the nodal lines of $\Delta_{\bm k}$;
this is clearly visible in the $k_z=0$ cut through the surface states shown
in~\fig{fig:boundState}(b). 

The SDOS is shown in~\fig{top_sdos}(a) for fixed
$\lambda=1$ and various values of $\Phi$. At $\Phi=0$ we recover the
SDOS for a 3D $d_{x^2-y^2}$ superconductor. For
non-zero $\Phi$ we observe that at low energies the SDOS is linear with a slope that
is controlled by the velocity of
the 2D linearly-dispersing surface states. The minimum energy of
the bulk gap is located at $E=0.5\Phi$, which is visible in the SDOS as
the point where the energy dependence becomes superlinear.
The linear slope of the low-energy SDOS does not change with
increasing $\Phi$, indicating that the velocity of the linear-dispersing
surface states is mainly determined by $\Delta$. We note that the finite SDOS
at $E=0$ is an artifact of the broadening parameter.

As shown in~\fig{top_sdos}(b), the SDOS is qualitatively different in the
topologically trivial ($\lambda<0.5$) and the nontrivial ($\lambda>0.5$)
regimes. In the former the low-energy SDOS is vanishing within the gap, while
in the latter there is the characteristic linear energy dependence. Note
that the sharp spike at $E=\Delta$ in the
$\lambda=0$ curve is the DOS peak associated with the purely $s$-wave
intraband gap.

\subsection{$(110)$ surface}

We now examine the bound states at the $(110)$ surface. Again setting
$\lambda=1$, we see that the condition $\text{sgn}(\Delta_{\bm{k}})\text{sgn}(\Delta_{\widetilde{\bm k}})=-1$ on the intraband potential is
satisfied for all ${\bm k}$. A zero-energy state thus only requires that ${\Phi}_{\bm k}$ vanishes for some
${\bm k}$ or that 
$\cos(\varphi_{\bm k} - \varphi_{\tilde{\bm k}}) = -1$. The former holds for
${\bm k}_1 = (k_{F},0,0)$ and $ {\bm k}_2 = (0,k_F,0)$, while the latter is
never realized.
Compared to the $(100)$ surface the roles of $\Delta_{\bm k}$ and $\Phi_{\bm{k}}$ are reversed: the zero-energy 
states are realized because ${\Phi}_{\bm k}$ vanishes at momenta where
$\Delta_{\bm k}$ has a sign change upon reflection. Furthermore, the position of
the zero-energy states do not change for $1>\lambda>0.5$, as the
sign change of $\Delta_{\bm k}$ at ${\bm k}_1$ and ${\bm k}_2$ survives while
the zeros of $\Phi_{\bm k}$ remain the same.

The energy of the $\lambda=1$
surface states is shown in units of the bulk gap
in~\fig{fig:boundState}(c). In contrast to the $(100)$ surface, most of the
interface states 
have energies differing significantly from $\Xi_{\bm k}$. This can be understood
as being due to the sign reversal upon reflection of $\Delta_{\bm k}$, which for
$\Phi_{\bm k}=0$ would give dispersionless zero-energy states for all ${\bm{k}}$.~\cite{kashiwaya_tanaka00,hu1994} In~\fig{fig:boundState}(d) we show a
$k_z=0$ cut of the interface states and bulk gap. 


\begin{figure}
\includegraphics[clip,width=1.0\columnwidth]{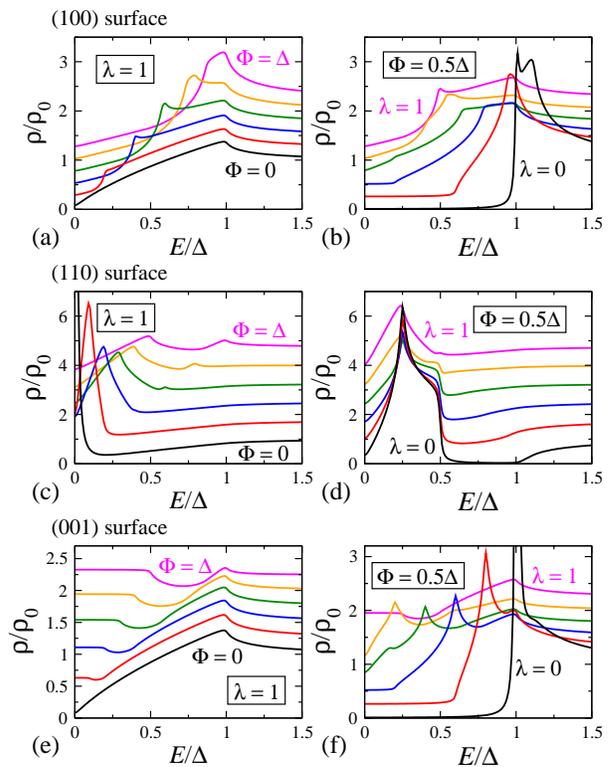}
\caption{\label{top_sdos} (color online) SDOS at the three surfaces as a  
function of the energy $E$, normalized by the constant SDOS $\rho_0$
of the normal state. $(100)$ surface: (a) fixed $\lambda=1$ 
and different values of $\Phi=0.2n\Delta$, and (b) fixed $\Phi=0.5\Delta$ and
different $\lambda=0.2n$, offset by $0.25n$, $n=0\ldots5$. $(110)$
surface: (c) fixed $\lambda=1$ 
and different values of $\Phi=0.2n\Delta$, and (d) fixed $\Phi=0.5\Delta$ and
different $\lambda=0.2n$, offset by $0.75n$, $n=0\ldots5$. $(001)$
surface: (e) fixed $\lambda=1$ 
and different values of $\Phi=0.2n\Delta$, and (f) fixed $\Phi=0.5\Delta$ and
different $\lambda=0.2n$, offset by $0.25n$, $n=0\ldots5$.}     
\end{figure}


In~\fig{top_sdos}(c) we show the change in the SDOS upon varying $\Phi$ at
fixed $\lambda=1$. Similarly to the $(100)$ surface, at $\Phi=0$ we recover
the results for a 3D $d_{xy}$ superconductor; note that the divergence of the
SDOS at $E=0$ is due to the zero-energy state for all ${\bm{k}}_\parallel$. As $\Phi$ is increased from zero, the SDOS becomes finite at
$E=0$ and increases linearly with $E$ up
to a maximum at the edge of the bulk gap. The low-energy linear
slope of the SDOS decreases with increasing $\Phi$, revealing that the
velocity of the Dirac states increases with $\Phi$. This is 
anticipated by the result that for this geometry the surface bound states
appear about the zeroes of $\Phi_{\bm k}$.

The change in the SDOS as $\lambda$ is tuned through the topological
transition is more subtle than for the $(100)$ surface. As can be seen 
in~\fig{top_sdos}(d), there is relatively little change in the low-energy
SDOS, with the main feature being that the peak at $E=0.5\Phi$
becomes sharper as $\lambda$ is decreased. At $\lambda<1$ we note a drop in
the SDOS at $E=0.5\Delta$; in the topologically nontrivial state, the
SDOS is finite on both sides of the drop, while in the trivial state it is
zero on the higher-energy side. That is, the large feature in the SDOS at
$E<0.5\Delta$ in the topologically trivial state is due to surface bound
states with non-zero energy.

\subsection{$(001)$ surface}

As there is no sign change of $\Delta_{\bm k}$ upon reflection from the
$(001)$ 
surface, the zero-energy states must be located along the nodal lines of
$\Delta_{\bm k}$. The interband pairing potential ${\Phi}_{\bm{k}}$ does not vanish for these values of ${\bm k}$, but 
the condition $\cos(\varphi_{\bm k} - \varphi_{\widetilde{\bm k}}) = -1$ is
fulfilled when the real part of ${\Phi}_{\bm k}$
vanishes, i.e. for ${\bm k}=(k_x,0, k_z)$ and $(0,k_y,k_z)$. For $\lambda=1$
this implies  
a single zero-energy bound state at ${\bm k}=(0,0,k_F)$. As for the $(100)$
surface, the origin of
this state is due to a zero in $\Delta_{\bm k}$ and the sign change 
of ${\Phi}_{\bm k}$ upon reflection. Upon reducing
$\lambda$, the 
single zero-energy state at $(0,0,k_F)$ splits into two zero-energy states at
the intersection of the plane $(0,k_y,k_z)$ with the nodal lines of
$\Delta_{\bm k}$. 

The energies of the surface state deviate most signficantly from the bulk
along the nodal lines of $\Delta_{\bm k}$, see~\fig{fig:boundState}(e). Unlike
the other two surfaces, the zero-energy  
state has quadratic dispersion at low energy, as shown by the cut along $k_x=0$
in~\fig{fig:boundState}(f). For $0.5<\lambda<1$, however, the two zero-energy
states have linear dispersion at low energies. 

The SDOS at $\lambda=1$ is qualitatively different to the other cases
because the 2D quadratic surface states contribute a constant DOS,
see~\fig{top_sdos}(e). The height of this constant region increases with
increasing $\Phi$, so that at $\Phi=\Delta$ the gap is completely filled. The
edge of the bulk gap at $E=0.5\Phi$ is signalled  
by the cusp feature. The constant SDOS within the bulk gap is only found at
$\lambda=1$: as shown in~\fig{top_sdos}(f), for $0.5<\lambda<1$ we find the
low-energy linear SDOS characteristic of linearly-dispersing zero-energy
states. In the topologically trivial state the SDOS is vanishing within the
gap.

\section{Conclusions and outlook}
\label{conclusion}

In this paper we have discussed the three-dimensional CI topological superconductor introduced in Ref.~\onlinecite{schnyderPRB08}.
We constructed a 
concrete realization of this topological phase in terms of a two-band Bogoliubov-de Gennes Hamiltonian 
with unconventional inter- and intraband pairing potentials. 
This lattice Hamiltonian is just one example of a wider class of models that all share the same topological properties. Quite generally, one is free
to add arbitrary small deformations to the Hamiltonian without changing its topological characteristics, as long as the perturbations
do not close the bulk superconducting gap. 
While we do not know in which specific material
the considered tight-binding Hamiltonian could be realized,
 it is a convenient canonical model that gives valuable insight into interesting properties shared by general CI topological superconductors.
In the presence of inversion symmetry, the topological characteristics of this two-band superconductor are fully determined by the momentum dependence of the gap functions along the  Fermi surface.
That is, the topological properties are independent of the electronic band structure away from the Fermi surface. 
We have demonstrated that the topological invariant $\nu$ (winding number) can be related to the sign reversal of the intraband gap 
between vortices in the interband gap with opposite winding number. This simple criterion could be used in the search for CI topological superconductors in real materials. Our results suggest to consider time-reversal invariant systems with orbital degrees of freedom, i.e., multi-band superconductors.

The CI topological superconductor has exotic Andreev bound states at its surface or at an interface with a normal metal. 
These gapless modes  are due to the bulk topological invariant $\nu$, which cannot change as long as the superconductor remains fully gapped in the bulk. We have used quasiclassical scattering theory to study the energy spectrum of these Andreev bound states for various surface orientations.
An important measurement technique to observe Andreev bound states in unconventional superconductors is scanning tunneling spectroscopy.
We therefore computed the surface density of states and demonstrated that the presence of Andreev bound states leads to pronounced anomalies  at low energies in the scanning tunneling spectra. These features provide key experimental signatures of the nontrivial topological character of the system. 

Furthermore, it would be interesting to investigate the effects of the
topological nontriviality of the superconductor  
on other experimental probes, such as tunneling conductance or Josephson
current. In particular, since our model lends itself to the
application of quasiclassical scattering techniques,  one could examine
the proximity effects in a junction involving a CI topological
 superconductor and a normal metal or a ferromagnet, for example, or
 alternatively examine vortex structures.
We leave these interesting questions for future work.

\acknowledgments

The authors thank S.\ Ryu, B.\ B\'eri, A.\ Ludwig, M.\ Sigrist, and G.\ Khaliullin for discussions. 
A.P.S.\ acknowledges the hospitality of the Max-Planck-Institut PKS Dresden, where part of this paper was
written.


\begin{thebibliography}{99}


\bibitem{kaneMelea08a} 
C.\ L.\ Kane and E.\ J.\ Mele, Phys. Rev. Lett. \textbf{95}, 226801 (2005). 

\bibitem{kaneMele08b}
C.\ L.\ Kane and E.\ J.\ Mele, Phys. Rev. Lett. \textbf{95}, 146802 (2005). 

\bibitem{bernevig06}
B.\ Bernevig, T.\ Hughes, and S.-C.\ Zhang, Science \textbf{314}, 1757 (2006). 

\bibitem{konig07}
M.\ K\"onig, S.\ Wiedmann, C.\ Brune, A.\ Roth, H.\ Buhmann, L.\ Molenkamp, X.\ Qi, and S.\ Zhang, Science \textbf{318}, 766 (2007). 

\bibitem{fuKanePRL07}
L.\ Fu, C.\ L.\ Kane, and E.\ J.\ Mele, Phys. Rev. Lett. \textbf{98}, 106803 (2007). 

\bibitem{fuKanePRB07}
L.\ Fu and C.\ L.\ Kane, Phys. Rev. B \textbf{76}, 045302 (2007). 

\bibitem{royPRB09}
R.\ Roy, Phys. Rev. B \textbf{79}, 195322 (2009).

\bibitem{mooreBalentsPRB07}
J.\ E.\ Moore and L.\ Balents, Phys. Rev. B \textbf{75}, 121306 (2007).

\bibitem{hsiehNature08}
D.\ Hsieh, D.\ Qian, L.\ Wray, Y.\ Xia, Y.\ Hor, R.\ Cava, and M.\ Hasan, Nature \textbf{452}, 970 (2008).

\bibitem{satoPRB06}
M.\ Sato, Phys.\ Rev.\ B \textbf{73}, 214502 (2006).

\bibitem{schnyderPRB08}
A.\ P.\ Schnyder, S.\ Ryu, A.\ Furusaki, and A.\ W.\ W.\ Ludwig, Phys.\ Rev.\ B \textbf{78} 195125 (2008).

\bibitem{royArxiv08}
R.\ Roy, arXiv:0803.2868 (unpublished). 

\bibitem{qiZhangPRB08}X.-L.\ Qi, T. L.\ Hughes, and S.-C.\ Zhang, Phys. Rev. B \textbf{78}, 195424 (2008).

\bibitem{schnyderPRL09}
A.\ P.\ Schnyder, S.\ Ryu, and A.\ W.\ W.\ Ludwig, Phys.\ Rev.\ Lett.\ \textbf{102}, 196804 (2009).

\bibitem{AIPconfProc}
A.\ P.\ Schnyder, S.\ Ryu, A.\ Furusaki, and A.\ W.\ W.\ Ludwig, AIP Conf.\ Proc.\ \textbf{1134}, 10 (2009).

\bibitem{kitaev09}
A.\ Y.\ Kitaev, AIP Conf.\ Proc.\ \textbf{1134}, 22 (2009).

\bibitem{qiHughesPRL09}
X.-L.\ Qi, T.\ L.\ Hughes, S.\ Raghu, and S.-C.\ Zhang, Phys. Rev. Lett. \textbf{102}, 187001 (2009).

\bibitem{satoPRB09}
M.\ Sato, Phys.\ Rev.\ B \textbf{79}, 214526 (2009).

\bibitem{satoFujiPRB09}
M.\ Sato and S.\ Fujimoto, Phys.\ Rev.\ B \textbf{79}, 094504 (2009).

\bibitem{hosur2010}
P.\ Hosur, S.\ Ryu, and A.\ Vishwanath, Phys. Rev. B \textbf{81}, 045120 (2010).

\bibitem{qiZhangPRB10}
X.\-L.\ Qi, T. L.\ Hughes, and S.-C.\ Zhang, Phys. Rev. B \textbf{81}, 134508 (2010).

\bibitem{schnyderNJP10}
S.\ Ryu, A.\ P.\ Schnyder, A.\ Furusaki, and A.\ W.\ W.\ Ludwig, New Journal of Physics \textbf{12}  065010 (2010).

\bibitem{readGreen}
N.\ Read and D.\ Green, Phys.\ Rev.\ B \textbf{61} 10267 (2000).

\bibitem{zirnbauerMathPhys96}
M.\ R.\ Zirnbauer, J. Math. Phys. \textbf{37}, 4986 (1996). 

\bibitem{altlandZirnbauer97}
A.\ Altland and M.\ R.\ Zirnbauer, Phys. Rev. B \textbf{55}, 1142 (1997).

\bibitem{bruderPRB90}
C.\ Bruder, Phys. Rev. B \textbf{41}, 4017 (1990). 

\bibitem{hu1994}
C.\ R.\ Hu, Phys. Rev. Lett. \textbf{72}, 1526 (1994).

\bibitem{tanaka_kashiwaya95}
Y.\ Tanaka and S.\ Kashiwaya, Phys. Rev. Lett. \textbf{74}, 3451 (1995). 

\bibitem{kashiwaya_tanaka96}
S.\ Kashiwaya, Y.\ Tanaka, M.\ Koyanagi, and K.\ Kajimura, Phys. Rev. B \textbf{53}, 2667 (1996).

\bibitem{kashiwaya_tanaka00}
S.\ Kashiwaya and Y.\ Tanaka, Rep.\ Prog.\ Phys.\ \textbf{63}, 1641 (2000). 

\bibitem{eschrigIniotakisReview}
M.\ Eschrig, C.\ Iniotakis, and Y.\ Tanaka, arXiv:1001.2486v1 (unpublished).

\bibitem{footnote1}
We note that the gap functions \eqref{IntraPair} and \eqref{InterPair} break the cubic symmetry of the
normal state Hamiltonian $h(\bm{k})$. A small symmetry-breaking term can be
included in $h ( \bm{k} )$, however, without changing the topological properties of
the system, and so we have neglected this in our analysis in the
interests of simplicity.

\bibitem{footnotePHS}
It is important not to confuse the particle-hole symmetry of the Bogoliubov-de Gennes Hamiltonian, Eq.~\eqref{def ham CI},
with the electron-hole symmetry of the normal state band structure, Eq.~\eqref{NSdispersion}, at half filling (i.e., $\mu_j = 0$).
The latter symmetry is irrelevant for the topological properties and does not play any role in our analysis.



\bibitem{beriPRB2010} B.\ Beri, Phys.\ Rev.\ B \textbf{81}, 134515 (2010).

\bibitem{murakamiPRB2008}
S.\ Murakami and S.\ I.\ Kuga, Phys.\ Rev.\ B \textbf{78}, 165313 (2008).




\end{thebibliography}
\end{document}